\documentclass{article}

 \title{The fundamental limit on the rate of quantum dynamics: the unified
bound is tight}
 \usepackage[paper=letterpaper,dvips]{geometry} \input pack9.sty
 \usepackage{amsmath,amsfonts,graphicx,url}
 \usepackage[superscript]{cite}

 \makeatletter
 \renewcommand{\@biblabel}[1]{$^{#1}$\!\!}
 \makeatother

 \input smallbib		

\makeatletter
\long\def\proofbox#1{\gdef\@proofbox{#1}}
\proofbox{\small{\tt\@ifundefined{sourcepath}{}{\sourcepath/}\jobname}\\%
\ifx\UNDEF\Version\else\quad v.\Version, \fi\today}

\def\proofref#1{\proofbox{\small{\tt#1\par
	[edited by Tom Toffoli for personal use]\par
	{\tt\sourcepath/\jobname}, 
	\number\month/\number\day/\number\year
	\par}}}

 \def\affil#1{\\{\small\sl#1\par}}
 \long\def\author#1{\gdef\@author{#1}}
 \author{Tommaso Toffoli ({\tt tt\char"40bu.edu})\affil{Electrical and
Computer Engineering, Boston University, MA 02215}}

 \long\def\abstract#1{\gdef\@abstract{#1}}
 \abstract{}

\long\def\@firstoftwo#1#2{#1}
\long\def\@secondoftwo#1#2{#2}
\def\@ifundefined#1{%
  \expandafter\ifx\csname#1\endcsname\relax
    \expandafter\@firstoftwo
  \else
    \expandafter\@secondoftwo
  \fi}
 \@ifundefined{leftfrac}{\def\leftfrac{.32}}{}
 \@ifundefined{ritefrac}{\def\ritefrac{.68}}{}

\def\@maketitle{\newpage\noindent\leavevmode
  \begin{minipage}[t]{\leftfrac\textwidth}
    \hrule height0pt
    \@proofbox
  \end{minipage}\hfil
 \begin{minipage}[t]{\ritefrac\textwidth}
    \hrule height0pt
    \raggedleft
    \LARGE\@title\par
    \vskip4pt
    \large\@author
  \end{minipage}
  \vskip8pt
  \ifx\@abstract\@empty\else{\vskip.5em\leftskip1.25in\parskip4pt\small\@abstract\par\vskip.5em}\fi
  \noindent
  \rule{\textwidth}{0.4pt}
  \vskip16pt}

\makeatother

 \def\leftfrac{.20}
 \def\ritefrac{.80}

\def\ifundefined#1{\expandafter\ifx\csname#1\endcsname\relax}

 \makeatletter
\def\code{\@code\frenchspacing\@vobeyspaces\verbatim@start}
\def\@code{\the\every@verbatim
 \par\noindent
  \@beginparpenalty \predisplaypenalty
  \leftskip=.4em\rightskip\z@
  \parindent\z@\parfillskip\@flushglue\parskip\z@
  \@@par
  \def\par{%
    \if@tempswa
      \leavevmode\null\@@par\penalty\interlinepenalty
    \else
      \@tempswatrue
      \ifhmode\@@par\penalty\interlinepenalty\fi
    \fi}%
  \def\@noitemerr{\@warning{No verbatim text}}%
  \obeylines
  \verbatim@font
  \let\do\@makeother \dospecials
  \everypar \expandafter{\the\everypar \unpenalty}}

 \makeatother

 \makeatletter

 \DeclareRobustCommand\em
        {\@nomath\em \ifdim \fontdimen\@ne\font >\z@
                       \upshape \else \slshape \fi}

\def\@begintheorem#1#2{\sl \trivlist \item[\hskip \labelsep{\bf #1\ #2}]}
\def\@opargbegintheorem#1#2#3{\sl \trivlist
     \item[\hskip \labelsep{\bf #1\ #2\ (#3)}]}

 \newcommand{\ie}{i.e.,}
 
 \newcommand{\eg}{e.g.,}

 \mathchardef\BY="0202

 \def\@empty{}
 \newcommand{\asin}[2][]{{
    \def\t@mp{#1}%
    \def\@cite##1##2{\marginpar{\hfil{\footnotesize$
    \ifx\t@mp\@empty\text{##2}\else\frac{\text{##1}}{\text{##2}}\fi$}\hfil}}%
\cite[#1]{#2}}}

 \def\pages#1{}

 \newcommand{\sectlabel}[1]{\label{sect:#1}}
 \newcommand{\footlabel}[1]{\label{foot:#1}}
 \newcommand{\eqlabel}[1]{\label{eq:#1}}
 \newcommand{\figlabel}[1]{\label{fig:#1}}
 
 \newcommand{\itmlabel}[1]{\label{itm:#1}}

 \newcommand{\Chapt}[2][]{\def\t@mp{#1}%
\chapter{#2} \ifx\t@mp\@empty\else\sectlabel{#1}\fi}
 \newcommand{\Sect}[2][]{\def\t@mp{#1}%
\section{#2} \ifx\t@mp\@empty\else\sectlabel{#1}\fi}
 \newcommand{\Subsect}[2][]{\def\t@mp{#1}%
\subsection{#2} \ifx\t@mp\@empty\else\sectlabel{#1}\fi}
 \newcommand{\Foot}[2][]{\def\t@mp{#1}%
\footnote{#2} \ifx\t@mp\@empty\else\footlabel{#1}\fi}
 \newcommand{\Itm}[2][]{\def\t@mp{#1}%
\item \ifx\t@mp\@empty\else\itmlabel{#1}\fi#2}
 \newcommand{\Eq}[2][]{\def\t@mp{#1}%
\begin{equation}#2\ifx\t@mp\@empty\notag\else\eqlabel{#1}\fi\end{equation}}
 \newcommand{\Eqaligned}[2][]{\def\t@mp{#1}%
\begin{equation}\begin{aligned}#2\end{aligned}
\ifx\t@mp\@empty\notag\else\eqlabel{#1}\fi
\end{equation}}
 \newcommand{\Eqmultline}[2][]{\def\t@mp{#1}%
\begin{multline}#2\ifx\t@mp\@empty\notag\else\eqlabel{#1}\fi
\end{multline}}
 \newcommand{\Eqgathered}[2][]{\def\t@mp{#1}%
\begin{equation}\begin{gathered}#2\end{gathered}
\ifx\t@mp\@empty\notag\else\eqlabel{#1}\fi
\end{equation}}

  \newcommand{\sect}[1]{\S\ref{sect:#1}}      

 \newcommand{\eq}[1]{(\ref{eq:#1})}	
 \newcommand{\fig}[1]{Fig.\,\ref{fig:#1}}

 \def\thlabel#1{\label{th:#1}}
 \def\theor#1{Theor.\,\ref{th:#1}}

 \newcommand{\Theor}[2][]{\def\t@mp{#1}%
\begin{theorem}#2\ifx\t@mp\@empty\else\thlabel{#1}\fi\end{theorem}}
 \newcommand{\Lemma}[2][]{\def\t@mp{#1}%
\begin{lemm}#2\ifx\t@mp\@empty\else\thlabel{#1}\fi\end{lemm}}
 \newcommand{\Defi}[2][]{\def\t@mp{#1}%
\begin{definit}#2\ifx\t@mp\@empty\else\thlabel{#1}\fi\end{definit}}
 \newcommand{\Prop}[2][]{\def\t@mp{#1}%
\begin{proposition}#2\ifx\t@mp\@empty\else\thlabel{#1}\fi\end{proposition}}

 \long\def\endsubsection#1{\smallskip\hbox to\hsize{\leaders\hrule\hfill\ \sect{#1}}\medskip}

  \def\@arabic#1{\number #1} 

\long\def\@makecaption#1#2{
	\vskip\abovecaptionskip
	\sbox\@tempboxa{{\small #1: #2}}%
	\ifdim\wd\@tempboxa>\hsize
	    {\small #1: #2\par}
	\else
	   \global\@minipagefalse
	   \hbox to\hsize{\hfil\box\@tempboxa\hfil}
	\fi
	\vskip\belowcaptionskip}

\def\figstrut#1{\hbox to\linewidth{\vrule height#1\hfill}}

\newcommand{\Fig}[3][]{
\begin{figure}[!htb]
 \centering{\leavevmode#2}%
 \caption{#3}
 \figlabel{#1}
\end{figure}                 }

\def\cstrip#1{\setbox0=\hbox{$#1$}\kern-.5\wd0\lower2pt\box0}
\def\rstrip#1{\setbox0=\hbox{$#1$}\kern-\wd0\lower2pt\box0}
\def\lstrip#1{\setbox0=\hbox{$#1$}\lower2pt\box0}
\def\tstrip#1{\setbox0=\hbox{$#1$}\kern-.5\wd0\lower\ht0\box0}
\def\bstrip#1{\setbox0=\hbox{$#1$}\kern-.5\wd0\raise\ht0\box0}
\def\Lstrip#1{\setbox0=\hbox{$\mskip2mu#1$}\lower2pt\box0}

\def\idpad{\thinspace}
\def\id{\idpad\begingroup \tt \let\do\@makeother \dospecials 
          \@ifstar{\@sid}{\@id}}
\def\@sid#1{\def\@tempa ##1#1{##1\endgroup\idpad}\@tempa}
\def\@id{\obeyspaces \frenchspacing \@sid}

\makeatother

 \input smallbib

 \newtheorem{lemm}{Lemma}
 \newtheorem{theorem}{Theorem}
 \def\theor#1{Theorem\,\ref{th:#1}}	

 \newtheorem{proposition}{Proposition}
 \newenvironment{proof}{\par{\sl Proof.}\quad}{\vrule height6pt width6pt depth0pt\par\medskip}

 \def\bra#1{\langle#1|}
 \def\ket#1{|#1\rangle}
 \def\braket#1#2{\langle#1|#2\rangle}
 \def\Re{\mathrm{Re}\,}
 \def\Im{\mathrm{Im}\,}
 \let\epsilon\varepsilon
 \def\Emax{E_{\text{max}}}

 \proofbox{\small {\tt \jobname}, \today}

 \author{Lev B. Levitin and Tommaso Toffoli}
 \title{The fundamental limit on the rate of quantum dynamics: the unified
bound is tight}
\begin{document}

\maketitle

{\small

{\bf \noindent The question of how fast a quantum state can evolve has
attracted a considerable attention in connection with quantum measurement,
metrology, and information processing. Since only orthogonal states can be
unambiguously distinguished, a transition from a state to an orthogonal one can
be taken as the elementary step of a computational process\cite{levitin82}.
Therefore, such a transition can be interpreted as the operation of ``flipping
a qubit'', and the number of orthogonal states visited by the system per unit
time can be viewed as the maximum rate of operation.

A lower bound on the orthogonalization time, based on the energy spread $\Delta
E$, was found by Mandelstam and Tamm\cite{mandelstam45}. Another bound, based
on the average energy $E$, was established by Margolus and
Levitin\cite{margolus98}. The bounds coincide, and can be exactly attained by
certain initial states if $\Delta E=E$. However, the problem remained open of
what the situation is when $\Delta E\neq E$.

Here we consider the unified bound that takes into account both $\Delta E$ and
$E$. We prove that there exist no initial states that saturate the bound if
$\Delta E\neq E$. However, the bound remains tight: for any given values of
$\Delta E$ and $E$, there exists a one-parameter family of initial states that
can approach the bound arbitrarily close when the parameter approaches its
limit value.  The relation between the largest energy level, the average
energy, and the orthogonalization time is also discussed. These results
establish the fundamental quantum limit on the rate of operation of any
information-processing system.}

\medskip

\noindent Starting with the classical result of Mandelstam and
Tamm\cite{mandelstam45}, it was later shown by Fleming\cite{fleming73}, Anandan
and Aharonov\cite{anandan90}, and Vaidman\cite{vaidman92} that the minimum time
$\tau$ required for arriving to an orthogonal state is bounded by
  \Eq[bnd1]{\tau\geq h/4\Delta E,}
 where $(\Delta
E)^2=\langle\psi|H^2|\psi\rangle-(\langle\psi|H|\psi\rangle)^2$, $H$ is the
Hamiltonian, and $|\psi\rangle$ the wavefunction of the system. A different
bound was obtained in \cite{margolus98}, namely,
  \Eq[bnd2]{\tau\geq h/4E.}
 Here, $E=\langle\psi|H|\psi\rangle$ is the quantum-mechanical average energy of the system (the energy of the ground state is taken to be zero). Both bounds
\eq{bnd1} and \eq{bnd2} are tight, and achieved for a quantum state such that 
$\Delta E=E$.

\medskip Since then, a vast literature has been devoted to various aspects of
this problem. In particular, inequality \eq{bnd2} has been proved for mixed
states and for composite systems both in separable and in entangled states
(\eg\ Giovannetti et al.\cite{giovannetti03,giovannetti03a}, Zander et
al.\cite{zander07}).  Bound \eq{bnd2} obtained for an isolated system has been
generalized to a system driven by an external Hamiltonian (a ``quantum gate'')
in \cite{levitin03operation_time,levitin05speed}. Various derivations of
\eq{bnd1} and \eq{bnd2} (\eg\ \cite{kosinski06,andrecut04,brody03}), bounds
based on energy-distribution moments\cite{zielinski06}, more general problems
of time-optimal quantum
evolution\cite{andrecut04,brody06,carlini06,luo04,pati99,pfeifer93,carlini07,borras07,soederholm99,uffink93},
and the ultimate limits of computation\cite{lloyd00,lloyd02} have been
considered.

However, what remained unnoticed is the paradoxical situation of the
existence of two bounds based on two different characteristics of the quantum
state, seemingly independent of one another. Since the average energy $E$ and
the energy uncertainty $\Delta E$ play the most determinative role in quantum
evolution, it is important to have a unified bound that would take into account
both of these characteristics.

In all known cases where bounds \eq{bnd1} and \eq{bnd2} can be exactly
attained, the ratio $\alpha=\frac{\Delta E}E$ equals 1. A question arises:
what happens if $\alpha\neq1$? Some authors just assumed without justification
that the minimum orthogonalization time is
 \Eq[bnd6]{
  \tau_{\text{min}} =\max\Big(\frac h{4E},\frac h{4\Delta E}\Big)
  =\frac h{2(E+\Delta E-|E-\Delta E|}.
 }
 In fact, the situation is not so simple. Bound \eq{bnd6}, indeed, can only
be achieved for $\alpha=1$. However, this bound remains tight when
$\alpha\neq1$ as well, though in this case it is only asymptotically
attainable.

\Theor[saturate]{Under the assumption that the smallest (ground) energy of a
quantum system is zero,
 \begin{enumerate}
  \item The only state that attains bound \eq{bnd1} is the
two-level state
 \Eq[diag]{
  \ket{\psi} = \tfrac1{\sqrt2}(\ket{\psi_0}+\ket{\psi_1}),
 }
 where $H\ket{\psi_k }= kE_1$ for $k=0,1$;
 \item The only state that attains bound \eq{bnd2} is likewise state
\eq{diag}.
 \end{enumerate}
  State \eq{diag} is unique up to degeneracy of the energy level $E_1$ and
arbitrary phase factors for $\ket{\psi_0}$ and $\ket{\psi_1}$.
 }

 \begin{proof}
 To prove the first statement we shall use the trigonometric inequality
 \Eq[trig]{
 \cos x \geq 1 - \frac4{\pi^2} x\sin x - \frac2{\pi^2}x^2,
 }
 which is valid for any real $x$. Note that \eq{trig} turns into an equality
iff $x=0$ or $x=\pm\pi$. 

Let the initial state be
 \Eq[init]{
 	\ket{\psi(0)}= \sum_{n=0}^\infty c_n\ket{E_n},
 }
 where the $\ket{E_n}$ are energy eigenstates of the system and
$\sum_{n=0}^\infty |c_n|^2=1$. Then
 \Eqaligned[sumsum]{
 |S(t)|^2 &= |\braket{\psi(0)}{\psi(t)}|^2
 \\ &= \sum_{n,n'=0}^\infty |c_n|^2|c_{n'}|^2 e^{-i\tfrac{E_n-E_{n'}}{\hbar/t}}
 \\ &=  \sum_{n,n'=0}^\infty |c_n|^2|c_{n'}|^2 \cos\frac{E_n-E_{n'}}{\hbar/t}.
 }

 Using inequality \eq{trig}, we obtain
 \Eqaligned[sob]{
 |S(t)|^2 &\geq 1-\frac4{\pi^2}\sum_{n,n'=0}^\infty |c_n|^2|c_{n'}|^2
  \dfrac{E_n-E_{n'}}{\hbar/t} \sin\dfrac{E_n-E_{n'}}{\hbar/t}
 \\&\phantom{=\ 1}- \frac2{\pi^2}\sum_{n,n'=0}^\infty|c_n|^2|c_{n'}|^2 \Big(\frac{E_n-E_{n'}}{\hbar/t}\Big)^2
 \\&= 1+\frac{4t}{\pi^2}\frac{d\,|S(t)|^2}{dt}-\frac1{\pi^2}\Big(\frac{\Delta E}{\hbar/2t}\Big)^2.
 }

 Since $|S(t)|^2\geq0$, it follows that $\frac{d\,|S(t)|^2}{dt}=0$ whenever
$S(t)=0$. Thus, at a time $\tau$ such that $S(\tau)=0$, the second term in
\eq{sob} vanishes, and we obtain
 \Eq[vanish]{
  0 \geq 1 - \frac{4\tau^2}{\pi^2 \hbar^2}(\Delta E)^2,
 }
 which yields inequality \eq{bnd1}; this is just another way to derive that
bound. However, for \eq{vanish} to turn into an equality, it is necessary that
inequality \eq{trig} turn into an equality for every term of the double
summation \eq{sumsum}. Hence,
 \Eq[or]{
  \text{either}\quad
  x_{nn'} = \frac{E_n-E_{n'}}{\hbar/\tau} = 0\quad\text{or}\quad
  x_{nn'} = \frac{E_n-E_{n'}}{\hbar/\tau} = \pm\pi
 }
 for all $n,n'$ such that $c_n\neq0, c_{n'}\neq0$. It follows that, to
attain bound \eq{bnd1}, $\ket{\psi(0)}$ must be a superposition of only two
energy eigenstates with energies $E_0=0$ and $E_1$.
 
To prove the second statement, we repeat briefly the derivation given in
\cite{margolus98}. This time we use the trigonometric inequality
 \Eq[trig1]{
  \cos x \geq1 -\frac2\pi(x+\sin x),
 }
 valid for all $x\geq0$. Again, \eq{trig1} turns into an inequality only for
$x=0$ or $x=\pi$.

Then
 \Eqaligned[reim]{
 \Re S(t)&=\sum_{n=0}^\infty |c_n|^2\cos\frac{E_n t}\hbar\\
         &\geq\sum_{n=0}^\infty |c_n|^2
    \Big[1-\frac2\pi \Big(\frac{E_nt}\hbar+\sin\frac{E_nt}\hbar)\Big]
 \\  &= 1 - \frac{2Et}{\pi\hbar}+ \frac2\pi \Im S(t).
 }

 At time $\tau$, $\Re S(\tau)=\Im S(\tau)=0$.
 Hence $0\ge 1-\frac{2Et}{\pi\hbar}$, which results in bound
\eq{bnd2}. However, to actually attain this bound, inequality \eq{trig1} must
become an equality for every term of the sum \eq{reim}; that is, for any $n$,
 \Eq{
  \text{either}\quad x_n =\frac{E_nt}\hbar=0\quad\text{or}\quad x_n=\frac{E_nt}\hbar=\pi;
 }
 This is possible iff $\ket{\psi(0)}$ has the form \eq{diag}.

Let us show now that no mixed state can attain bound \eq{bnd6}. Consider an
initial mixed state density matrix $\rho(0)$ with spectral decomposition
 \Eq[spectral]{
 \rho(0) = \sum_i \lambda_i\ket{\psi^{(i)}(0)}\bra{\psi^{(i)}(0)},
 }
 \noindent where $\lambda_i>0$, $\sum_i \lambda_i=1$, and
$\braket{\psi^{(i)}(0)}{\psi^{(j)}(0)}=\delta_{ij}$. At time $t$, the density matrix
becomes
 $$
 \rho(t) = \sum_i \lambda_i\ket{\psi^{(i)}(t)}\bra{\psi^{(i)}(t)}.
 $$
 \noindent Hence, if $\tau$ is the orthogonalization time, then 
 \Eq[trace]{
 \text{Tr}[\rho(0)\rho(\tau)]=\sum_i\sum_j\lambda_i\lambda_j
 |\braket{\psi^{(i)}(0)}{\psi^{(j)}(t)}|^2=0.
 }
 \noindent Note that all terms in the above sum are non-negative, and
therefore all of them must be zero to satisfy \eq{trace}. As shown above,
terms $\braket{\psi^{(i)}(0)}{\psi^{(i)}(\tau)}=0$, for $\tau$ given by
\eq{bnd6}, iff each $\ket{\psi^{(i)}(0)}$ has the form \eq{diag}.

Consider two functions from spectral decomposition \eq{spectral}
  \Eqaligned[spectral_dec]{
  \ket{\psi^{(i)}(0)} &= \tfrac1{\sqrt2}\big(\ket{\psi_0}+\ket{\psi^{(i)}_1(0)}\big)\quad\text{and}\\
  \ket{\psi^{(j)}(0)} &= \tfrac1{\sqrt2}\big(\ket{\psi_0}+a\ket{\psi^{(i)}_1(0)}+
					b\ket{\psi^{(j)}_1(0)}\big),
 }
 \noindent where $\braket{\psi^{(i)}(0)}{\psi^{(j)}(0)}=0$ and $|a|^2+|b|^2=1$
 (the zero-energy state is nondegenerate). Since
 $$
 \braket{\psi^{(i)}(0)}{\psi^{(j)}(0)}=\tfrac12\big(\braket{\psi_0}{\psi_0}+
 a\braket{\psi_1^{(i)}(0)}{\psi_1^{(i)}(0)}\big)=0,
 $$
 \noindent  it follows that $a=-1$ and $\ket{\psi^{(j)}(0)}= 
\frac1{\sqrt2}\big(\ket{\psi(0)}-\ket{\psi^{(i)}_1(0)}\big)$. But then
 $$
 \braket{\psi^{(i)}(0)}{\psi^{(j)}(\tau)}=
 \tfrac12\big(\braket{\psi_0}{\psi_0}-
 \braket{\psi_1^{(i)}(0)}{\psi_1^{(i)}(\tau)}\big)=\tfrac12\neq0.
 $$

 Hence, equality \eq{trace} cannot be satisfied for $\tau$ given by
\eq{bnd6}. Thus, this bound is not attainable by a mixed state.
 \end{proof}

For pure states, results of \theor{saturate} also follow from the analyses
presented in \cite{kosinski06} and \cite{zielinski06}.

\medskip

\theor{saturate} shows that bounds \eq{bnd1} and \eq{bnd2} can only be
attained by a state for which $\alpha = \frac{\Delta E}E = 1$.  Such a state
is unique up to degeneracy of the energy level $E_1$ and arbitrary phase
factors for $\ket{\psi_0}$ and $\ket{\psi_1}$. It follows from that theorem
that there exists no initial state
 $\ket{\psi(0)}$ with $\alpha\neq1$ that would attain bound \eq{bnd6}, \ie\
no state can evolve into an orthogonal state in the minimum time given by
\eq{bnd1} or \eq{bnd2}. The question to be answered is how close it is
possible to approach the unified bound \eq{bnd6}. Let us rewrite \eq{bnd6} in
a different form,
 \Eq[alternate]{
 \tau_{\text{min}} = \max\Big(\frac h{4E},\frac h{4\Delta E}\Big) = 
 \frac {h(1+e^{|\ln\alpha|})} {4E(1+\alpha)}.
 }

\Theor[exist]{For any $\alpha\neq1$ and any $\epsilon>0$, there exists a state
$\ket{\psi(0)}$ such that $\braket{\psi(0)}{\psi(\tau)}=0$ at time
 \Eq[time]{
  \tau \leq \frac{h(1+e^{|\ln\alpha|})}{4E(1+\alpha)} (1+\epsilon).
 }
 }

 \begin{proof} We will show that in both cases, $\alpha<1$ and $\alpha>1$,
there exist families of initial states that approach limit \eq{alternate}
arbitrarily close.

1. Let $\alpha<1$. Consider a family of states
 \Eq[family]{
  \ket{\psi(0)} = c_0\ket{0}+c_1\ket{E_1}+c_2\ket{E_2}.
 }
 Denote $|c_i|^2=p_i$, and introduce dimensionless variables
 \Eq[dimless]{
  x_i(t)=2\pi E_i\frac th   ,\quad i=0,1,2
 }
 (the rotation angles of the state vectors $\ket{E_i}$). Then
 $S(\tau) = \braket{\psi(0)}{\psi(\tau)} = 0$
 iff $x_i(\tau)=x_i$ such that
 \Eq[p1p2]{
  p_1\sin x_1+p_2\sin x_2=0
 }
 and
 \Eq[p0p1p2]{
  p_0+p_1\cos x_1+p_2 \cos x_2 = 0.
 }
 Of course,
 \Eq[norm]{
  p_0+p_1+p_2 = 1.
 }
 Let $0<p_0<<1$ be a parameter of the family of states \eq{family} (one
cannot set $p_0=0$, since that would change the ground energy level). When
$p_0$ is small, the values of $x_1$ and $x_2$ differ almost exactly by
$\pi$. Let
 \Eq[delta]{
  x_2=\pi+x_1-\delta\sin x_1,\quad \text{with}\ \delta<<1.
 }
 Then, from  \eq{p1p2},
 \Eq[delta_cos]{
  p_1 = p_2[1-\delta\cos x_1+\text{O}(\delta^2)].
 }
 Substituting \eq{delta_cos} into \eq{p0p1p2} and \eq{norm} we obtain
 \Eqaligned[p1p2p0]{
  p_1 &= \tfrac12 - \tfrac\delta4(1+\cos x_1)+\text{O}(\delta^2),\\
  p_2 &= \tfrac12 - \tfrac\delta4(1-\cos x_1)+\text{O}(\delta^2),\\
  p_0 &= \tfrac\delta2+\text{O}(\delta^2).
 }
 Using \eq{delta} and \eq{p1p2p0}, we can calculate the average value
 $\langle x\rangle=\tfrac{2\pi\tau}h E$, namely,
 \Eqaligned[xmean]{
  \langle x\rangle & = p_1x_1+p_2x_2\\
 &=\tfrac\pi2+x_1-\tfrac\delta4[2x_1-2\sin x_1+\pi(1-\cos x_1)]+\text{O}(\delta^2).
 }

 The standard deviation $\Delta x=\frac{2\pi\tau}h \Delta E$ is obtained as
 \Eqaligned[sqrt]{
  \Delta x &= \sqrt{p_1x_1^2+p_2x_2^2-(p_1x_1+p_2x_2)^2}\\
    &=\tfrac\pi2 +\tfrac\delta{2\pi}[x_1(\pi+x_1)-\pi\sin x_1]+\text{O}(\delta^2).
 }
 On the other hand,
 \Eq[===]{
  \alpha = \frac{\Delta E}E = \frac{\Delta x}{\langle x\rangle}
	= \frac\pi{\pi+2x_1} + \text{O}(\delta).
 }
 Hence,
 \Eq[1/alpha]{
  x_1 = \tfrac\pi2\big(\tfrac1\alpha-1\big)+\text{O}(\delta  ).
 }
 Substituting \eq{1/alpha} into \eq{sqrt} and taking into account the last
expression from \eq{p1p2p0} yields
 \Eq[yields]{
  \Delta x=\frac\pi2+p_0\Big[\tfrac\pi4\big(\tfrac1{\alpha^2}-1\big)-\sin\tfrac\pi2\big(\tfrac1\alpha-1\big)\Big]
           +\text{O}(p_0^2),
 }
 which results in
 \Eqaligned[results]{
  \tau &= \frac{h\Delta x}{2\pi\Delta E}\\
       &= \frac h{4\Delta E}\Big[1+\frac{p_0}2\Big(\frac1{\alpha^2}-1-\frac4\pi \sin2\pi\big(\tfrac1\alpha-1\big)\Big)+\text{O}(p_0^2)\Big].
 }
 Finally, choosing
 \Eq{
  p_0<2\epsilon\Big[\frac1{\alpha^2}-1-\frac2\pi\sin\tfrac\pi2\big(\frac1\alpha-1\big)\Big]^{-1},
 }
 we obtain
 \Eq[finally]{
  \frac h{4\Delta E}<\tau\leq\frac h{4\Delta E}(1+\epsilon).
 }

 2. Let $\alpha>1$. Consider a family of states
 \Eq[famk]{
  \ket{\psi(0)}=c_0\ket{0}+c_1\ket{E_1}+c_{2k+1}\ket{E_{2k+1}},
 }
 where
 \Eqaligned[where]{
 &k=1,2,\dots;\ E_{2k+1}=(2k+1)E_1;\ |c_0|^2 = p_0=\tfrac12;\\
 &|c_1|^2=p_1=\tfrac12\big(1-\tfrac\beta{k^2}\big);\ |c_{2k+1}|^2 =
p_{2k+1}=\tfrac\beta{2k^2}.
 }
 Using the dimensionless variables \eq{dimless}, it is readily seen from
\eq{famk} and \eq{where} that $S(\tau)=0$ at the least time $\tau$ for which
 \Eq[least]{
  x_1(\tau)=x_1=\tfrac{2\pi\tau}h E_1=\pi\ \ \text{and}\ \ 
  x_{2k+1} =\tfrac{2\pi\tau}h(2k+1)E_1=\pi(2k+1).
 }
 Then
 \Eq[xmeanbis]{
 \langle x\rangle=p_1x_1+p_{2k+1}x_{2k+1}=\tfrac\pi2\big(1+\tfrac{2\beta}k\big)
 }
 and
 \Eqaligned[sqrtbis]{
 \Delta x &=\sqrt{p_1x_1^2+p_{2k+1}x_{2k+1}^2-(p_1x_1+p_{2k+1}x_{2k+1})^2}\\
          &=\tfrac\pi2\big(1+4\beta+\tfrac{2\beta}k)+\text{O}\big(\tfrac1{k^2}\big).
 }
 Hence, $\alpha=\frac{\Delta x}{\langle x\rangle}=1+4\beta+\text{O}\big(\tfrac1k\big)$ and
 \Eq[hence]{
  \beta= \frac{\alpha-1}4+\text{O}\Big(\frac1k\Big).
 }
 Since $\langle x\rangle=\frac{2\pi\tau}h E$, it follows from \eq{sqrtbis} and
\eq{hence} that
 \Eq[follows]{
 \tau=\frac h{4E}\Big[1+\frac{\alpha-1}{2k}+\text{O}\Big(\frac1{k^2}\Big)\Big].
 }
 Thus, choosing a sufficiently large $k>\frac{\alpha-1}{2\epsilon}$
guarantees that
 \Eq[proof2end]{
 \frac h{4E}<\tau\leq\frac h{4E}(1+\epsilon).
 }
 This completes the proof.
 \end{proof}

As graphically illustrated in \fig{keel}, \theor{exist} means that there are
families of states that approach equality in \eq{bnd6} for a given value of
$\alpha$, in the limit as a certain parameter approaches zero (for
$\alpha<1$) or infinity (for $\alpha>1$), but no state yields a strict
equality in \eq{bnd6} for $\alpha\neq1$. It turns out that bound \eq{bnd1} is
tight whenever $\alpha=\frac{\Delta E}E\leq1$, while \eq{bnd2} is tight
whenever $\alpha\geq1$.

\Fig[keel]{\def\XMAG{1.25}\unitlength\XMAG bp\small
 \def\HSIZ{210}\def\VSIZ{111}\def\XOFS{-106}\def\YOFS{-6}
 \begin{picture}(\HSIZ,\VSIZ)(\XOFS,\YOFS)
  \put(-100,0){\line(1,0){200}}
  \put(-100,0){\line(0,1){105}}\put(100,0){\line(0,1){105}}
  \put(0,0){\line(0,1){2}}
  \put(-50,0){\line(0,1){2}} \put(50,0){\line(0,1){2}}

  \put(0,-6){\cstrip{1}}
  \put(-100,-6){\cstrip{0}}\put(100,-6){\cstrip{\infty}}

  \put(-50,-6){\cstrip{\frac13}}\put(50,-6){\cstrip{3}}

  \multiput(-100,20)(0,20){5}{\line(1,0){2}}
  \put(-106, 0){\rlap{0}}\put(-106,20){\rlap{1}}
  \put(-106,40){\rlap{2}}\put(-106,60){\rlap{3}}
  \put(-106,80){\rlap{4}}\put(-106,100){\rlap{5}}

  \put(0,20){\circle*{2}}
  \put(83,10){\cstrip{\alpha=\tfrac{\Delta E}E}}
  \put(75,3){\vector(1,0){10}}

  \put(-94,100){\lstrip{2\tau(E+\Delta E)/h}}
  \put(-97,95){\vector(0,1){10}}

  \put(-74,80){\lstrip{\longleftarrow p_0}}\put(-77,80){\rstrip{0}}
  \put( 74,80){\rstrip{k\longrightarrow}}  \put( 77,80){\lstrip{\infty}}

  \put(-100,0){\includegraphics[scale=\XMAG]{fig/tau.eps}}
 \end{picture}%
 }
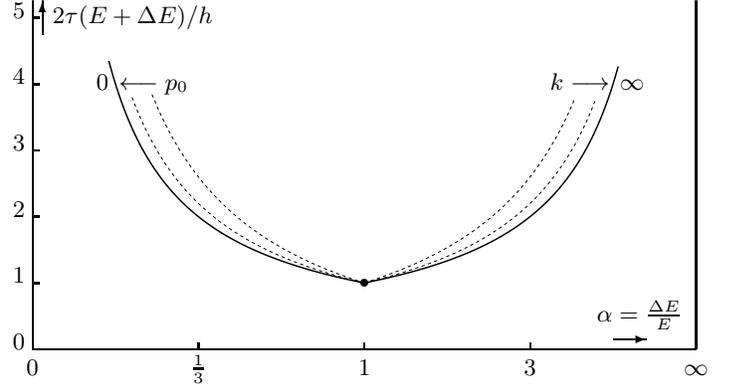
 {The solid line shows bound \eq{alternate}; the dotted lines---parametrized,
respectively, by $p_0\to0$ and $k\to\infty$---show successive approximations
to this bound, corresponding to initial states from families \eq{family} and
\eq{famk}.}

\bigskip

Another interesting question is the relationship between the maximum energy
eigenvalue $\Emax$ that contributes to $\ket{\psi(0)}$, the
orthogonalization time, $\tau$, and the average energy $E$. Note that
determining the minimum $\tau$ for a given $E$ is equivalent to determining
the minimum $E$ for a given $\tau$. The next theorem provides simple but
useful results.

\Theor[last]{Let $\tau$ be the minimum time such that
 $S(\tau)=\braket{\psi(0)}{\psi(\tau)}=0$. Then there exist a state
$\ket{\psi(0)}$ and a set of energy eigenvalues $\{E_n\}$ of all energy
eigenfunctions that contribute to $\ket{\psi_0}$ such that
 \Eq[emax]{
	\Emax<\frac h\tau
 }
 and
 \Eq[emax1]{
	\frac{\Emax}4 \leq E \leq \frac{\Emax}2.
 }
 }

 \begin{proof} Let 
 \Eq[Stau]{
	S(\tau)=\sum_{n=0}^\infty|c_n|^2 e^{-2\pi i E_n\tau/h} = 0.
 }
 Suppose $E_k\geq\frac h\tau$. Then $E'_k=E_k-\frac h\tau\geq0$ and
$e^{-2\pi i E_k\tau/h} =  e^{-2\pi i E'_k\tau/h}$. Hence, replacing $E_k$ by
$E'_k$ in \eq{Stau} will not affect the equality. Thus, the same ortogonalization time $\tau$ can be achieved with smaller average energy $E'=E-|c_k|^2h/\tau$.
This proves inequality \eq{emax}.

Now, let $\Emax$ be the largest energy of the energy eigenfunctions
that contribute to $\ket{\psi(0)}$, and let the average energy be $E^{(1)}$.
Obviously, the validity of \eq{Stau} will not be affected if we replace all
energy levels $E_n$ by $\Emax-E_n$. Then the average energy will
become $E^{(2)}=\Emax-E^{(1)}$. Since $E$ can be chosen as
$\min(E^{(1)},E^{(2)})$, this proves that $E\leq \Emax/2$. Also,
substituting \eq{bnd2} into \eq{emax} yields $\Emax/4\leq E$.
Thus, \eq{emax1} is proved.
 \end{proof}
 } 

\noindent\theor{last} allows us to restrict our attention to states that
satisfy \eq{emax} and \eq{emax1} for further analysis of the orthogonalization
time $\tau$.

\Theor[onemore]{Let $\Emax$ be the maximum energy of all energy
eigenfunctions that contribute to $\ket{\psi(0)}$. Then the minimum
orthogonalization time $\tau$ obeys the inequality
 \Eq[ineq]{
 \tau\geq\frac h{2\Emax},
 }
 \noindent and equality is attained iff $\ket{\psi(0)}$ has the form
\eq{diag} with $E_1=\Emax$.}

\begin{proof} Let 
 \Eq[S_of_tau]{
 S(\tau)=\braket{\psi(0)}{\psi(\tau)}=\sum_{n=0}^m|c_n|^2 e^{-2\pi iE_n/h}=0,
 }
 \noindent where $E_m=\Emax$. Then, if $\tau<\frac h{2\Emax}$, all terms in the
above sum except the zeroth have strictly positive imaginary parts---which
violates equality \eq{S_of_tau}. Similarly, if $\tau=\frac h{2\Emax}$, then
\eq{S_of_tau} is satisfied iff $\ket{\psi(0)}$ is of form \eq{diag} with
$E_1=\Emax$.
 \end{proof}
 \noindent The equality case in \eq{ineq} had been considered in \cite{brody03}.

\medskip

Some authors (\eg\ \cite{brody03}) prefer to write expressions \eq{bnd2} and
\eq{ineq} as $\tau \geq h/4(E-E_0)$ and $\tau \geq h/2(E_{\text{max}}-E_0)$,
arguing that a constant shift of the energy spectrum affects only the overall
time-dependent phase of the wavefunction. In our opinion, however, it is just
this ``freedom of shift'' that allows one to set to zero the energy of the
ground state even if the smallest eigenvalue of the Hamiltonian is not zero
(just as in the case of the hydrogen atom).

\medskip

The authors wish to thank Prof.\ D.~Brody (Imperial College, London) for a useful
discussion.

\end{document}